\documentclass[submission,copyright,creativecommons]{eptcs}
\providecommand{\event}{Third Workshop on Agents and Robots for reliable Engineered Autonomy} 

\usepackage{iftex}
\usepackage{graphicx}
\usepackage{tabularx}
\usepackage{makecell}
\usepackage{caption}
\usepackage{microtype}
\ifpdf
  \usepackage{underscore}         
  \usepackage[T1]{fontenc}        
\else
  \usepackage{breakurl}           
\fi

\title{ORTAC+ : A User Friendly Domain Specific Language for Multi-Agent Mission Planning}
\author{Caroline Bonhomme
\institute{Safran Electronics and Defense\\ Massy, France}
\institute{ONERA\\
Toulouse, France}
\email{caroline.bonhomme@safrangroup.com}
\and
Jean-Louis Dufour
\institute{Safran Electronics and Defense\\
Massy, France}
\email{jean-louis.dufour@safrangroup.com}
}
\def\titlerunning{ORTAC+ : A Domain Specific Language for Multi-Agent Mission Planning}
\def\authorrunning{C. Bonhomme, J-L. Dufour}
\begin{document}
\maketitle

\begin{abstract}
A tactical military unit is a complex system composed of many agents such as infantry, robots, or
drones. Given a mission, an automated planner can find an optimal plan. Therefore, the mission itself
must be modeled. The problem is that languages like PDDL are too low-level to be usable by the
end-user: an officer in the field. We present ORTAC+, a language and a planning tool designed for
this end-user. Its main objective is to allow a natural modeling of the mission, to minimize the risk of bad modeling, and thus obtain reliable plans. The language offers high-level constructs specifically
designed to describe tactical missions, but at the same time has clear semantics allowing a translation
to PDDL, to take advantage of state-of-the-art planners.
\end{abstract}

\section{Introduction}
\vspace{-2mm}
 
A tactical military ground unit is a complex system composed of many agents, such as infantry sections, armored units, ground robots, or flying drones. A typical mission lasts one day and is a mix of displacements on a road network of 20 km between given start and finish points and actions. We deal only with the reconnaissance action of a route or an area when an enemy presence is suspected. For us, the reconnaissance action is equivalent to a speed reduction of the progression on a route and exploration of an area. The progression and exploration should be done more safely, thus it is slower. It could be seen as the first visit to a location. This brings us to a variant of the classic Multi-Agent Path Finding (MAPF) problem \cite{stern2019multi}.  Of course, if we consider moves as actions, this is also a multi-agent planning problem \cite{10.1145/3128584}, but classification as a MAPF problem is more accurate.

Given a mission, if it is described in a programming language, an automated planner can find an optimal plan. Therefore the mission itself must be modeled. The automated planning community has developed the PDDL language in order to specify high level planning problems \cite{haslum2019introduction}. However, the specification of the mission should be done by an officer in the field who is not an expert in automated planning. Languages like PDDL are too low-level to be usable by the defined end-user who is not an expert in planning. 
This is why we present ORTAC+. Its main objective is to allow a natural modeling of the mission, so as to minimize the risk of errors on the modeling of the desired mission, and thus obtain reliable plans.
The language is composed by high-level constructs specifically designed to facilitate the specification of tactical missions, with a clear semantic allowing a translation to PDDL, to take advantage of state-of-the-art planners.
Planning is done offline, and the plans are deterministic, with no contingency. The default assumption is that the enemy, when suspected, is in fact not present. Otherwise, if there is contact with the enemy, the execution of the plan is de facto interrupted in the field, and the officer must replan with modified objectives.

\indent We proceed by presenting related work of the paper in section 2. The domain-specific language ORTAC+ is presented in section 3, the operational context in section 4, and the conclusion in section 5.

\section{Related Work}
\vspace{-2mm}
ORTAC stands for "Optimal Resource and Technical Action Control". It is a multi-agent planning language and tool combining constraint-solving techniques with advanced search strategy to deconflict single-agent plans \cite{guettier2015design}. At the language level, ORTAC is innovative in the constraints of coordination between units introducing the notion of "support" explained later \cite{guettier2015design}. The proper term for that is "cross-schedule dependencies" \cite{korsah2013comprehensive}, typically in the case of "coalition formation."

PDDL is the standard planning language used commonly by the automated planning community. One of the purposes of PDDL was to benchmark planning algorithms.
Since its creation in 1998 for the International Planning Competition \cite{aips98}, the language evolved to model more complex planning problems as temporal or numerical planning \cite{haslum2019introduction}. The Competition of Distributed and Multiagent Planners \cite{vstolba2016competition} benchmarks some centralized and decentralized algorithms with an extension of PDDL for multiagent planning as a reference language. 
ORTAC and PDDL are powerful tools for modeling high-level missions and finding a corresponding plan, nevertheless, they cannot be used by an end-user.

In recent years, an effort on designing domain specific languages (DSL) for non-expert in robotics raised in the research community. Indeed, some of these languages lack clear semantics \cite{Dragule2021}.  Additionnaly, users are required to define the behavior of each robot. Consequently, the autonomy of the system is hindered. Theses systems lack planning capabilities, making it challenging to initiate the integration of planning tools. The research community uses PDDL as the standard task planing language. However, its integration into robotics remains limited. To remedy this issue,  efforts were done to integrate planning in robotics \cite{karpas2020automated}.

\section{ORTAC+}
\vspace{-2mm}

ORTAC+ is a tool that assists a military officer in planning a mission. It is a language that aims to be expressive with well-defined syntax and semantics. To do that, the language should allow the description of a given mission and be compatible with automated planners. A military mission is limited in space and time. The first version of ORTAC+ emphasizes the spatial aspects of the mission. The temporal aspects will be added in future works.

\subsection{Problem Definition}
\vspace{-2mm}

The missions could be seen as a multi-agent planning problem where each agent collaborates to reach a global goal. However, we only consider agents movements. Thus, the model is a variant of a MAPF problem. A classical MAPF problem is represented with a set of agents with an initial position and a target position, in an undirected graph. The solution of a classical MAPF problem is a set of single-agent paths that does not collide \cite{stern2019multi}. ORTAC+ allows the description of this kind of problem but aims to be more general with the following variants: 
\begin{itemize}
  \setlength\itemsep{-2pt}

    \item  Targets are no longer assigned to each agent but to teams of agents, leading to the possibility of having less target positions than agents. Several agents could be assigned to the same target position but only one of them should reach it.
    \item Spatial constraints are added to the problem. The constraint can be forbidden or enforced vertex for specific agent. 
    \item Agents can be heterogeneous. While classical MAPF problems harnesses agents with similar characteristics, our problem differs in this aspect. 
    \item The position of an agent is not only on vertices but can be on edges too. The path of an agent is then a succession of vertices and edges.
\end{itemize}

\subsection{Modelisation}
\vspace{-2mm}

Our DSL describes the mission within a three axis representation:
\begin{itemize}
  \setlength\itemsep{-2pt}
    \item A geographical representation, corresponding to an undirected graph $G=(V,E)$, with $V$ the space of vertices which models geographic points and $E$ the space of edges that links one vertice to another. If there is an edge between two vertices, it means that there is a path between these two points, a road for example. 
    \item A resource representation, including the agents involved in the mission and their characteristics. Each element of the description is an object with attributes.
    \item An operational representation where the constraints and the goal will be defined.
\end{itemize}

This approach allows a structured and organized representation of critical information, enabling a comprehensive analysis of a complex mission. While the geographic and resource representations correspond to static meta-data available at the start of the mission, the operational representation corresponds to flexible operational and tactical elements from the mission, allowing the definition of the goal and mandatory waypoints. Furthermore, some components of the description are objects with attributes. The type of mission modeled by ORTAC+ deals with three types of objects: the nodes, the edges, and the agents.
By leveraging the three axes, we enable an in-depth portrayal of mission-critical information with a more expressive and accessible representation. It is worth noting that the agents involved in the missions are usually humans, but the tool aims to be compatible with robots.

During the conception phase, the graph is pre-defined or already given. The mission specification includes the available resources and the operational elements within the system with high-level predicates. Each predicate has a logical meaning and is a constraint for a planning algorithm. 

\subsubsection{Resources}
\vspace{-2mm}

The resources encompass the agents, their initial states, and their characteristics. The \textit{agent$\_$define(init, characteristics)} predicate instantiates the agent, creating the object agent with its initial position \textit{init} and its \textit{characteristics}. It is still possible to add attributes to the agent after its instantiation.

\subsubsection{Operational Predicates}
\vspace{-2mm}

This representation is the modular part of the modeling. Indeed, modeling a complex mission can lead to the absence of solutions. In this part the user can change some elements of the problem and relax some constraints to find a plan. The planning tool allocates a path for each agent to reach the goal, and the tactical interpretation of the mission remains to the user.

\begin{table*}[ht]
\centering
\begin{tabularx}{\textwidth}{|XXX|}
\hline
\textbf{Predicates} & \textbf{Description} & \textbf{Logic formula} \\
\hline
node$\_$goal(node, agent) & The location of agent should be node at the final state & $agent.loc(t_{final}) = node$ \\
\hline
node$\_$visit(node, agent) & Agent should visit node during execution & $\exists t \in [\![t_{initial};t_{final}]\!], \newline agent.loc(t) = node$ \\
\hline
edge$\_$visit(edge, agent) & Agent should visit edge during execution & $\exists t \in [\![t_{initial};t_{final}]\!],\newline agent.loc(t) = edge$ \\
\hline
node$\_$avoid(node, agent) & Agent should never visit node & $\forall t \in [\![t_{initial};t_{final}]\!] ,\newline agent.loc(t) \neq  node$ \\
\hline
edge$\_$avoid(edge, agent) & Agent should never visit edge & $\forall t \in [\![t_{initial};t_{final}]\!], \newline agent.loc(t) \neq  edge$ \\
\hline
node$\_$supported$\_$from(node$_1$, node$_2$) & An agent can visit node$_1$ only if another agent supports him in node$_2$ & $\forall t \in [\![t_{initial};t_{final}]\!],   \forall agent, \newline agent.loc(t) =  node_1 \Rightarrow \newline \exists  agent_2 \neq agent,\newline agent_2.loc(t) = node_2$ \\
\hline
\end{tabularx}
\caption{Existing predicates in ORTAC+ DSL. Other predicates can be constructed}
\label{tab:predicates}
    \vspace{-12pt}

\end{table*}

Table \ref{tab:predicates} presents the high-level predicates introducing the constraints. Each predicate has a description and a logical meaning. For example, the predicate \textit{node$\_$goal()} can allocate a node goal to an agent. The predicates aim to give a declarative specification.

Besides, the predicates can be generalised so that they operate on lists of objects and no longer individual objects. To facilitate modeling, when a "generalised" predicate is applied to an individual object, this is automatically replaced by the application on the singleton containing that object. So the preceding predicate is in fact syntactic sugar for:
\textit{node_goal([node],[agent])}.
The general case
\textit{node_goal(node_list,agent_list)}
has the following meaning:
\vspace{-2mm}
\begin{equation}
    \forall n \in node\_list, \exists a \in agent\_list, a\ is\ in\ n\ at\ t_{final}
\end{equation}

The $\forall\ n $ is in fact a general pattern to which all 'list-generalized' predicates conform, so as to facilitate their understanding: \textbf{the first list argument can always be "and-expanded"}, which means:

predicate([o1,o2], list2, ...) $\Leftrightarrow$ predicate(o1, list2, ...) \& predicate(o2, list2, ...)

Because of the multi-agent nature of the model, coordination constraints could be required. We are interested in the coordination presented in ORTAC \cite{guettier2015design}. The main coordination predicate uses quantitative time, but if we restrict ourselves to causality, it boils down to:
\textit{\textit{support(unit1,node1,unit2,node2)} }\\
which means: "when unit1 goes through node1 (if it goes there), unit2 must be in node2". For this to make sense, node1 and node2 must be close enough, and from a tactical point of view, it is the standard way to progress in dangerous areas. However, we are no longer specifying the mission, but have begun designing the plan. We will call ORTAC's description style "imperative", and ORTAC+ will add a new style that we will call "declarative":
\textit{node_supported_from(node1,node2)}
which means: "when an agent goes through node1, another agent must be in node2". 
This first addition to ORTAC will be called "anonymization", it is not a generic evolution because it is only suitable for certain predicates like \textit{support} or \textit{goal}.  And concerning goals, their "anonymization" corresponds to an interesting variant of MAPF initially called "permutation-invariant" \cite{Kloder2006} and today called "anonymous" \cite{ma2016optimal}.

\subsubsection{Ontology}
\vspace{-2mm}

Considering the specificity of our language for its domain, the proposal entails adding a knowledge base to augment its capability. This knowledge base is an ontology designed to help the mission specification. The characteristics of the agents is defined with the declarative predicate presented in the section 3.2.1.

It is possible to specify a constraint involving several agents. For example, considering a mission with two types of agents, an agent can be a "UAV" for unmanned aerial vehicle or a "UGV" for unmanned ground vehicle. This characteristics will be define as an attribute of the agents. Instead of writing: \textit{node$\_$goal(14, [agent1, agent2])}, it is possible to write \textit{node$\_$goal(14, "UGV")} if agent1 and agent2 have the attribute UGV, meaning at the end of the mission a "UGV" should be at the node 14.

Another example of the combined efficiency of objects and ontologies is: french infantry uses impressive vehicles called "VBCI" which need some place to maneuver. If the officer wants to ban the progression of these vehicles on narrow roads, he has to stipulate: \textit{edge_avoid("width < 10", "VBCI")} in spite of \textit{edge_avoid(list_of_edges_with_width_less_than_10, list_of_units_with_VBCI)}

Currently, the ontology is hierarchical. An attribute can be a leaf in knowledge tree. For instance, if agent1 is a UGV on wheels and agent2 is a UGV on caterpillars. There are the relations: a UGV with wheels is a UGV, and the UGV on caterpillars is a UGV. They have the same parent. The ontology allows to specify \textit{node$\_$goal(14, "UGV")} to involve agent1 and agent2 in the constraint. 

\section{Operational Context}
\vspace{-2mm}

ORTAC+ is built to specify military missions in order to assist an officer planning the mission. In the following part, an example of mission is described, then its representation in the proposed language.
\vspace{-2mm}
\subsection{Description of the Mission}
\vspace{-2mm}
\begin{figure*}[ht]
    \centering
    \begin{minipage}[t]{0.35\linewidth}
        \centering
        \includegraphics[width=\linewidth]{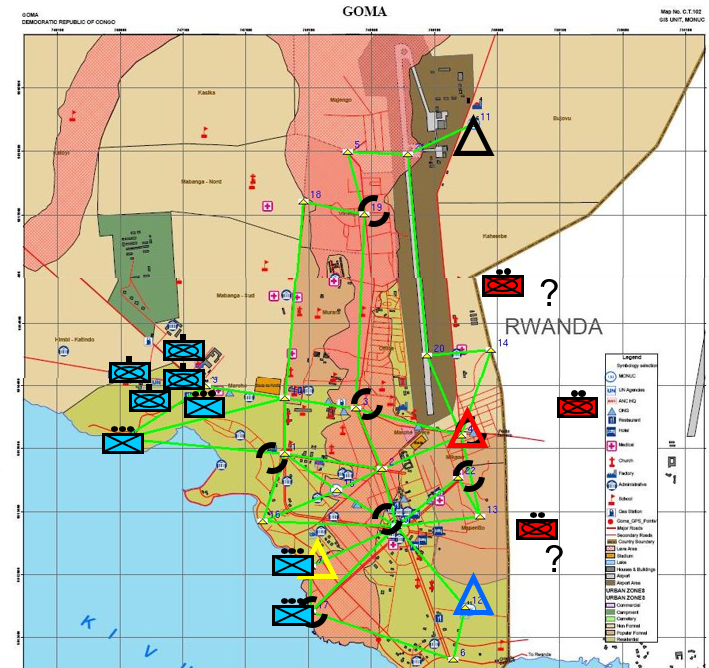}
        \caption{Cartography of the mission in Goma, Congo.}
        \hspace{1mm}
        \label{fig:mission}
    \end{minipage}%
    \hspace{0.1\linewidth}
    \begin{minipage}[t]{0.35\linewidth}
        \centering
        \includegraphics[width=\linewidth]{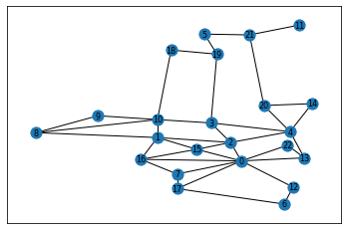}
        \caption{Geographical mission modelisation using graphs }
        \hspace{1mm}
        \label{fig:graph}
    \end{minipage}
\end{figure*}


\indent Congolese rebels, supported by elements from Rwanda, are located in the eastern region of Goma. The rebels raise a significant threat to the peace and security of the city. The mission focused on two specific objectives: securing the United Nation (UN) point at the border, between Congo and Rwanda, and preparing for the potential evacuation of civilians at the airport. With forces comprising 4 distinct military sections and 4 motorized companies positioned along the west cost of the city, this mission implies the recognition of crucial points and axes leading to the destination goals. \\
\textbf{Operational constraint.} Units follow specific rules of engagements during the execution of their tasks:
\begin{itemize}
  \setlength\itemsep{-2pt}
    \item Simultaneous coexistence of two distinct units at the same spatial coordinates is not allowed. 
    \item Any unit advancing on a main street or an intersection must be supported by another unit recently deployed to an adjacent position.
\end{itemize}
\noindent
\textbf{Tactical information.} The mission encompasses critical data as a foundation for modeling and planning:
\begin{itemize}
  \setlength\itemsep{-2pt}
    \item The airport: the airport is a crucial point and must be controlled during the mission, with a designated convoy of one company and one unit equipped with night vision glasses will be tasked to evacuate people from the airport.
    \item  Securing the UN point: The UN point is identified at the border, requiring strict security measures, with a priority to maintain control over 
    \item Support team: A support team will be positioned at the southern region of the operation area
\end{itemize}

\subsection{Mission specification in ORTAC+}
\vspace{-2mm}
\indent
\begin{figure}
    \centering
    \includegraphics[width=0.6\linewidth]{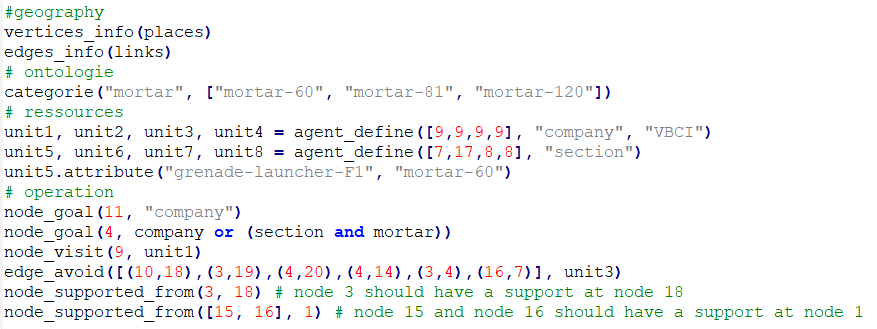}
    \caption{Specification of the "Secure the UN in Goma" mission in ORTAC+}
    \label{fig:code}
    \vspace{-12pt}

\end{figure}
An example of the mission specification in ORTAC+ is given in Figure \ref{fig:code}. The mission goal is the securing of the United Nations. The cartography of the mission is given in Figure \ref{fig:mission}. Friendly forces are blue rectangles at the west of Goma. Target positions are triangles at the east near to enemies in red rectangles. The black triangle correspond to the airport and the red one the UN. 
The 3 axis of representation (geography, resources and operations) appear clearly combined with the construction of a hierarchical ontology. 
Geography elements of the mission are supposed to be already given. It is represented in Figure \ref{fig:graph}
The Goma mission involves two types of agents: compagnies and sections. The predicates "\textit{agent$\_$define()}" is used to instantiate the units. At the initial state, the 4 mechanized compagnies are at the same node (9). This violates the constraint that only one unit can be at a node at the same time. That's why the notion of capacity is introduced for nodes and edges. By default the capacity of a node or an edge will be 1 but it can be changed for some exceptions. 
The method "\textit{attribute()}" allows the user to add some characteristics to the units.
Then, the high level predicates described earlier are used to add the constraints of the mission. The final state is defined with the predicate "\textit{node$\_$goal()}" with the help of the knowledge base. A compagny should be at the node 11, corresponding to the airport, at the final state. Other constraints are defined, for instance, the unit1 should visit the node 9 but avoid the edge (9,8). ORTAC+ introduces coordination constraints. If an enemy is suspected on a node or an edge, a unit cannot visit it without support. Therefore, the predicate "\textit{node$\_$supported$\_$from()}" model this coordination. In the presented example, it is specified in order to visit the node 3, there must be a support at the node 18.

\section{Conclusion}
\vspace{-2mm}
We have presented ORTAC+, a DSL that hopefully permits a military officer to specify a mission. Compared to its predecessor, ORTAC, the evolution can be summarized as follows:
\begin{enumerate} 
    \setlength\itemsep{-2pt}    
    \item predicates are anonymized when suitable, in particular for the declaration of goals or supports and operate on lists of homogeneous objects (agents, nodes, or edges),
    \item these lists can be described intentionally, with a propositional sub-language constraining the attributes of objects,
    \item these propositions, when they involve the "or" operator, can be shortened thanks to the declaration of the adequate ontology.
\end{enumerate}
As a result, ORTAC+ allows more compact and declarative models, which will be more easily written and checked by end-users. At the same time, we take care to keep semantics to allow a translation to PDDL, making it possible to rely on the existing planners. Current and future work explores 3 orthogonal directions:  new coordination constraints, quantitative time and resources, and compatibility with flying drones.

\newpage
\nocite{*}
\bibliographystyle{unsrturl}
\bibliography{generic}
\end{document}